# Binary Fingerprints at Fluctuation-Enhanced Sensing

**Hung-Chih Chang[1], Laszlo B. Kish[1,*], Maria D. King[2], and Chiman Kwan[3]**

1 Department of Electrical and Computer Engineering, Texas A&M University, College Station, TX 77843-3128, USA

2 Department of Mechanical Engineering, Texas A&M University, College Station, TX 77843- 3123, USA

3 Signal Processing, Inc., 13619 Valley Oak Circle, Rockville, MD 20850, USA

* Author to whom correspondence should be addressed: Laszlo@ece.tamu.edu

Abstract: We developed a simple way to generate binary patterns based on spectral slopes in different frequency ranges at fluctuation-enhanced sensing. Such patterns can be considered as binary "fingerprints" of odors. The method has experimentally been demonstrated with a commercial semiconducting metal oxide (Taguchi) sensor exposed to bacterial odors (*Escherichia coli* and *Anthrax*-surrogate *Bacillus subtilis*) and processing their stochastic signals. With a single Taguchi sensor, the situations of empty chamber, tryptic soy agar (TSA) medium, or TSA with bacteria could be distinguished with 100% reproducibility. The bacterium numbers were in the range of $2.5*10^4$ - $10^6$. To illustrate the relevance for ultra-low power consumption, we show that this new type of signal processing and pattern recognition task can be implemented by a simple analog circuitry and a few logic gates with total power consumption in the microWatts range.

## 1. Introduction

Bacterium detection and identification has an important role in medical, agricultural, environmental, defense, etc. applications. Analyzing their odor [1][2] has good prospects because of high speed, low cost, wide availability, good sensitivity and selectivity, while solid-state electronic noses [3-7] can be applied.

Recently, we have carried out an experimental study [8] with commercial Taguchi sensors to test the shape of the power density spectrum of the stochastic component of their signal as a pattern to recognize bacteria. The power density spectrum $S(f)$ of the spontaneous fluctuations of the sensor signal is one of the easiest and natural tools for Fluctuation-Enhanced Sensing (FES) of chemicals [9-20]. While it is reasonably simple to generate it from the measured data, it contains significant sensing information and it has been shown to enhance sensitivity by a factor of 300, or more [14,16]. It is also relatively straightforward to construct a theory to explain its behavior [18-20].

In the present paper, we show a new method to generate binary patterns from measured spectra, an ultra-low power implementation of such a system including a simple Boolean logic circuit as a microprocessor-free pattern recognizer, see Sections 2, 3 and 6, respectively.

In order to demonstrate the feasibility of the method and the nature of binary patterns, we conducted relevant experimental tests/evaluations where we have used some of the spectra published in paper [8] and spectra from new measurements.

## 2. Binary patterns for low power consumption

To achieve ultra-low power consumption, we must avoid the usage of microprocessors and extensive data processing. The sensor signal must be processed in the simplest possible way, presumably with analog circuitry, and the pattern recognition must be a deterministic process based on a few simple logic decisions.

Let us make the following notations: $\alpha_n$ (local slope) is the average local slope of the power density spectrum $S(f)$ in the $n$-th frequency sub-band and $\beta = \sum \alpha_n / N$ is the average of $\alpha_n$ over the entire measurement band, see Figure 1 as an illustration for logarithmically equidistant sub-band boundaries. The boundaries of sub-bands can be equidistant or any convenient settings. These quantities can easily be generated by a low number of operational amplifiers and filters, see Figures 2 and 3.

**Figure 1.** Illustration of $\alpha_n$ and $\beta$ for logarithmically equidistant sub-band boundaries. $S(f)$ is the power density spectrum of the fluctuations of the sensor signal.

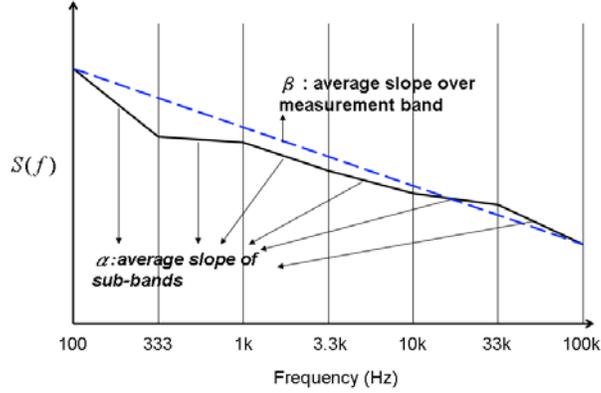

The deviation $\Delta_n$ of the local slope is defined for each sub-band as the difference between $\alpha_n$ and $\beta$ in the following equation

$$\alpha_n - \beta = \Delta_n \qquad (1)$$

The sign $\sigma_n$ of the local deviation $\Delta_n$ will be binary bit related to that sub-band:

$$\sigma_n = \text{signum}(\Delta_n) \qquad (2)$$

The quantity $\sigma_n$ is a binary pattern that indicates if $\alpha_n$ is larger or smaller than $\beta$ in the $n$-th sub-band of the spectrum. The advantage of the quantity $\sigma_n$ is that it provides a single bit information about the spectral pattern. In the case of $N$ non-overlapping frequency bands, the $\sigma_n$ ($n = 1,...,N$) quantities represent $N$ bit information obtained from a single sensor. Then a simple, deterministic, fast and low-power pattern recognizer can easily be constructed by applying a Boolean logic rule to identify/distinguish the particular spectral patterns with their relevant set of the $\sigma_n$ bits.

All these tasks can be realized without the use of power hungry devices such as microprocessors, other types of sequential logic, or analog-digital converters. In Section 3, we show a simple realization with analog circuitry of power consumption in the microWatt range and in Section 7, based on experimental patterns, a Boolean pattern recognition logic in the nanoWatt/picoWatt regime, see also Section 8.

## 3. An ultra-low-power realization of the scheme

The proposed system for ultra-low-power consumption, see Figure 2, includes three major parts: Sensor, Analog circuits to generate the binary patterns, and a Boolean Logic circuit. The system described here uses equidistant sub-band boundaries: *uniform bandwidths with non-overlapping sub-bands*.

**Figure 2.** Major building blocks of the low-power sensing system

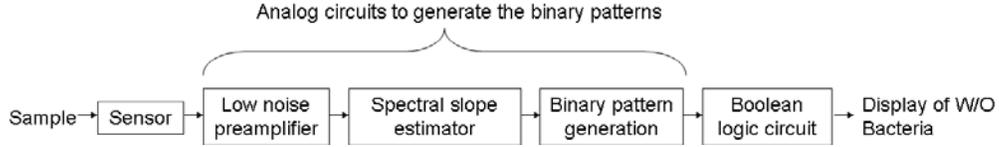

In Figure 3, the outline of the Analog Unit is shown. The preamplifier amplifies the stochastic component of sensor signal. The spectral slope estimator is the combination of a small number of amplifiers, filters and rectifiers. The binary patterns are generated by a set of comparators.

**Figure 3.** Details of the Analog Unit of the sensing system with 6 bit resolution shown in Figure 2, see the text for explanation.

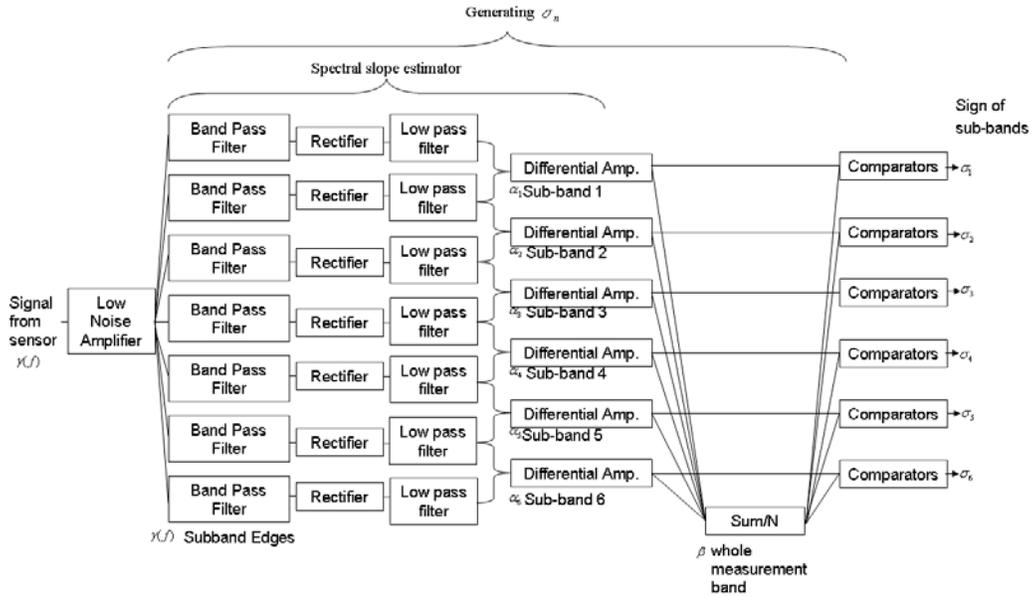

A simple realization of the analog electronics to estimate the local slope $\alpha_n$ and global slope $\beta$ can be seen in the Spectral slope estimator part of Figure 3. In this realization, for fingerprints of $N$ bit resolution, we will need $N+1$ sub-bands. To obtain the local slope $\alpha_n$ of each sub-band, first we rectify and then smooth the noise by a low-pass filter at the output of each band-pass filter in order to estimate the mean-square amplitude $\langle U_n^2(t) \rangle$ there. Then we estimate the average slope in the frequency band $\{(f_n - f_{n-1})/2 \; ; \; (f_n + f_{n+1})/2\}$ as:

$$\alpha_n = \frac{\langle U_n^2(t) \rangle - \langle U_{n-1}^2(t) \rangle}{(f_n - f_{n-1})^2} \qquad (3)$$

To estimate $\alpha_n$ from the mean-square voltages, see Equation 4, differential amplifiers are used, see Figure 3. To obtain the magnitude of $\beta = N^{-1} \sum \alpha_n$ the $\alpha_n$ values obtained above are averaged with an adder with amplification of 1/$N$. Finally, the comparators compare the magnitude of $\alpha_n$ and $\beta$ and generate the bit $\sigma_n$ for all sub-bands. The Boolean logic circuit will identify the pattern of bits $\sigma_n$ ($n = 1,...,N$).

## 4. Experiments with bacteria and heated semiconducting metal oxide sensors

To demonstrate the feasibility of the new method, the number of bits required for the identification, the simplicity, and the low power consumption of such a system, we carried out experiments with bacteria and commercial Taguchi sensors. Note, commercial Taguchi sensors must be heated thus they consume a lot of power, however, nanoparticle film alternatives are often much more sensitive and do not need heating during sensing [21,22].

Thus, whenever the new method described in this paper will be used for ultra-low power applications, nanoparticle sensors or similar room-temperature devices will be needed to fully utilize the low power consumption of the electronics.

*4.1. Sample preparation*

Here we briefly summarize the sample preparation steps. For more details, see [8]. As Gram negative pathogenic vegetative bacterium surrogate, mid-log phase cultures of *E. coli* K12 MG1655 (*E. coli* Genetic Resources at Yale CGSC, New Haven, NE) were grown in Luria Bertani (LB) medium [23] at 37°C. The cells were harvested at 2880 g for 9 minutes and resuspended in 5% Phosphate Buffer Saline (PBST, pH 7.4) to $10^9$ CFU/milliLiter concentration. Aliquots (100 microLiter) of the *E. coli* cell suspension were spread in appropriate dilution on Difco Tryptic Soy Agar (TSA) plates (Becton Dickinson Co., Sparks, MD), and incubated overnight at 37°C [23].

As Gram positive (*Anthrax*) surrogate, 50 mg of lyophilized *Bacillus atrophaeus* (aka *Bacillus globigii*, BG) (U.S. Army Edgewood Proving Ground, Edgewood, MD) was resuspended in 5 milliLiter of sterile deionized water and centrifuged at 2880 g for 9 min to remove traces of the culture medium. The supernatant was aspired and the pellet was resuspended in 10 milliLiter of sterile deionized water. Aliquots (100 microLiter) of the stock *Bacillus subtilis* were spread in appropriate dilution on TSA plates and incubated overnight at 30°C [23].

As reference, sterile TSA plates containing identical (27 milliLiter) amounts of the medium without bacteria were also prepared.

*4.2. Experimental setup [8]*

The data shown below were measured on a single commercial Taguchi sensor SP32 (Figaro Inc.) after the new sensor was preheated ("burned-in") in laboratory air for several days with the nominal heating voltage until its stochastic signal component (resistance fluctuations) developed a stable power density spectrum $S_r(f)$. In addition, to enhance the sensitivity and selectivity of the sensor, the *Sampling-and-hold* (SH) method, see [8,10], was used. At the SH protocol the sensor is heated for a short time, and then the heating (and gas flow) is turned off and, after the sensor has cooled down, the measurement is done [8,10].

**Figure 4.** Outline of the experimental setup.

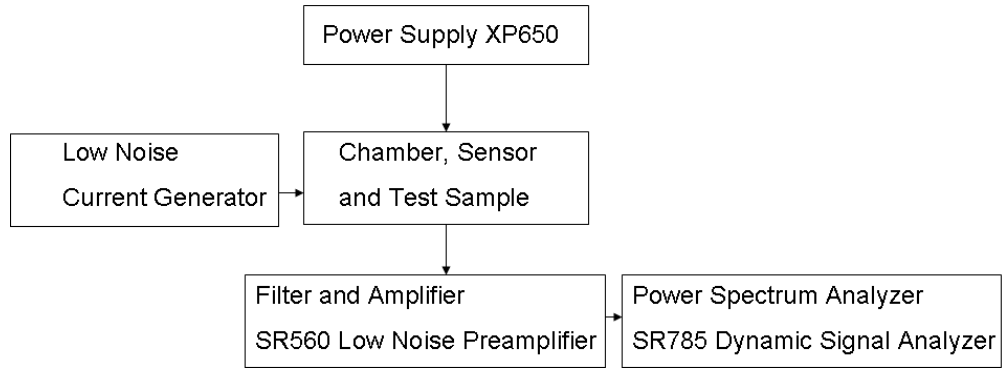

The measurement system [8] is shown in Figure 4. The sensor and the sample were in a grounded stainless steel sensor chamber of one Liter volume. The sample in a Petri-dish was located 5 cm below the sensor. A DC bias through the sensor converted its resistance fluctuations into voltage fluctuations that were amplified by a low-noise preamplifier and their power density spectra were evaluated by a dynamic signal analyzer in the frequency range of 100Hz~100kHz. The voltage spectra were transformed back to power density spectra $S_r(f)$ of the fluctuations of sensor resistance [8-10].

*4.3. Types of samples*

The situations tested in these experiments were: empty chamber; TSA; TSA + *E. coli* at different bacterium numbers; TSA + *Anthrax*-surrogate; finally TSA + *E. coli* + *Anthrax*-surrogate. Here empty means no sample; TSA represents the culture medium: tryptic soy agar; E. coli means the harmless laboratory strain MG1655 as a surrogate for the pathogenic vegetative bacteria *Escherichia coli*; and *Anthrax* stands for the *Anthrax*-surrogate bacterium, the spore forming *Bacillus subtilis*. The samples with maximal bacterium number had one million bacteria.

Between different sample measurements, the chamber was flushed by synthetic air for 3 minutes. To see the reproducibility of the spectra, the measurements with each sample were repeated at least twice.

**5. Binary pattern extracted from experiments**

In the figures below, for the sake of better visibility of the differences by the naked eye, the normalized power density spectrum $\gamma(f)$ is used:

$$\gamma(f) = \frac{f * S_r(f)}{R_s^2}, \qquad (4)$$

where $R_s$ is the actual sensor resistance. Note, for the binary patter generation describe in Section 2, both the original and the normalized spectra yield the same result.

For simplicity, we measured the average slopes in six sub-bands with logarithmically equidistant width by fittings of the $\gamma(f)$ plotted as a log-log plot by Origin software. The sub-bands were 100~333Hz for bit B1, 0.333~1kHz for bit B2, 1~3.3kHz for bit B3, 3.3~10kHz for bit B4, 10~33kHz for bit B5 and 33~100kHz for bit B6. The binary pattern used for driving the logic circuit is found to have the following characteristics.

**(i)** *Good reproducibility*: Examples are shown in Figure 5-8: measurement data obtained with independently prepared samples at different dates. The spectra in Figures 5-7 yield identical patterns shown in Figure 8.

**Figure 5.** Normalized power density spectra of the resistance fluctuations of the sensor SP32 measured in the sampling-and-hold [8, 10] working mode. Each sample had one million bacteria. The alias "*Anthrax*" stands for *Anthrax* surrogate *Bacillus subtilis*.

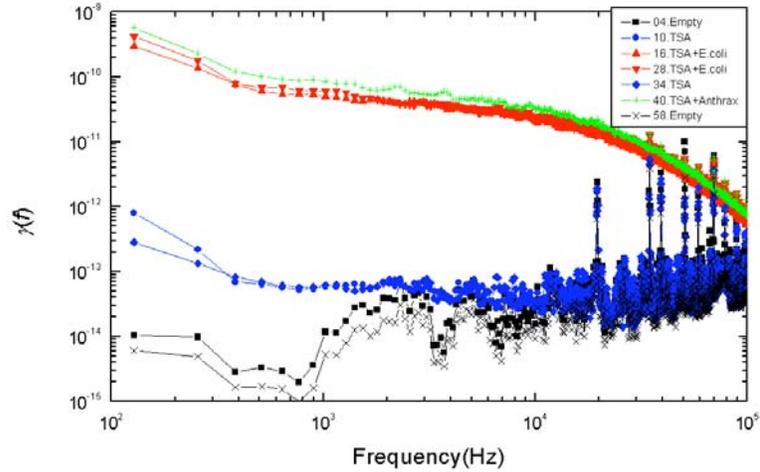

**Figure 6.** Reproducibility of the experimental data shown in Figures 5 with new samples.

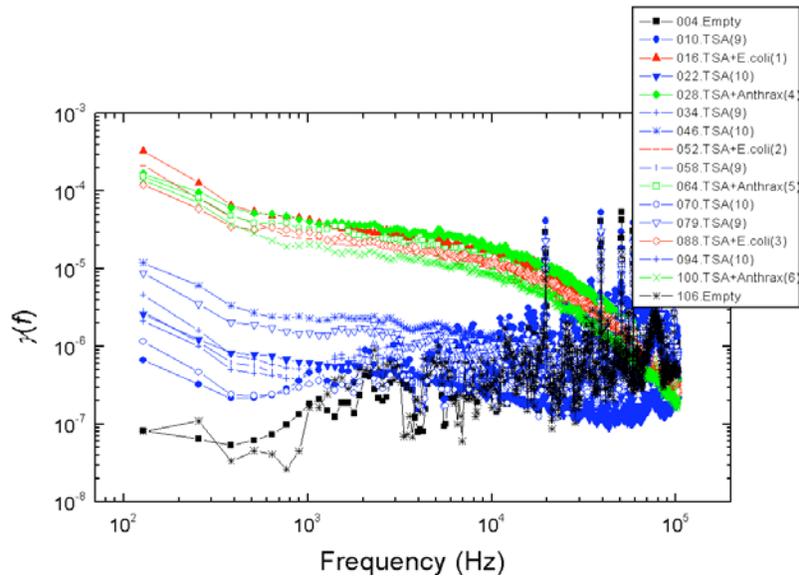

**Figure 7.** Reproducibility of the experimental data shown in Figures 5-6 with new samples.

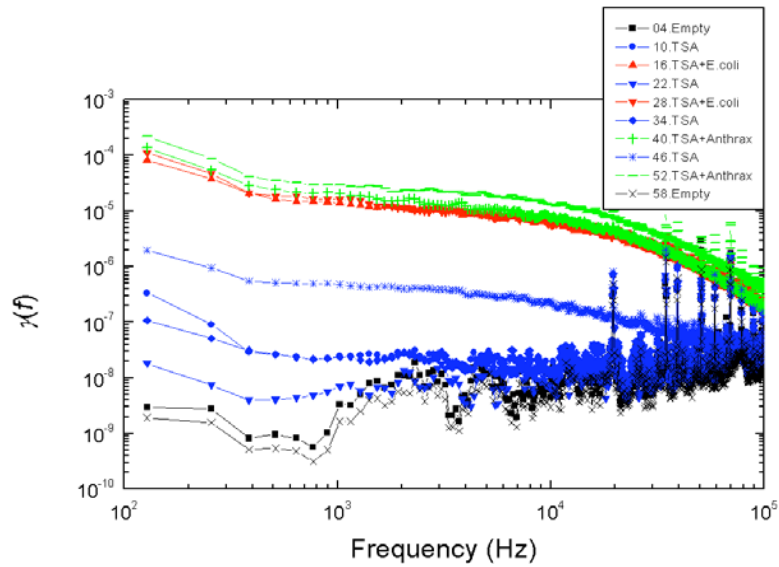

**Figure 8.** The spectra in Figures 5-7 yield the same 6-bits pattern.

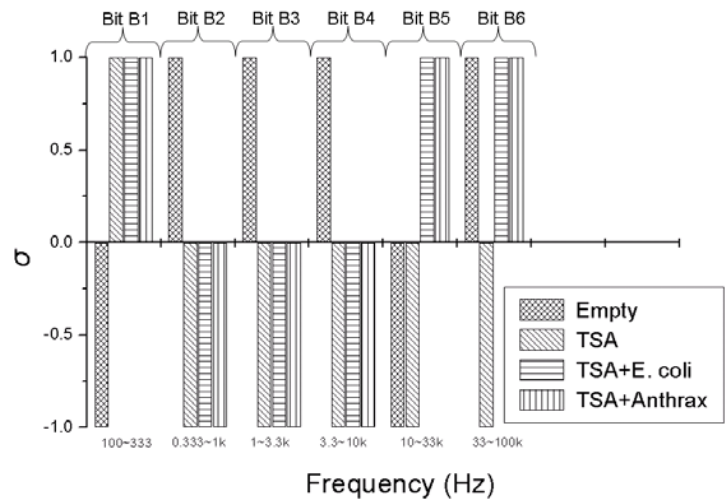

**(ii)** *Inability to differentiate between the two types of bacteria*: the applied sensor and the simple 6 bit pattern generation we used for these tests were unable to differentiate between the two bacteria, while they were able to differentiate between all the other cases (empty, TSA, bacteria). This fact originates from the particular settings of pattern generation because the differences between spectra with different bacteria could be distinguished by naked eye. However, we find this situation satisfactory because our goal was not to present a fully featured/optimized system but to show how much can be achieved with just a simple, ad-hoc, demo version of a 6 bits system.

**(iii)** *Robustness against variations of the bacterium number*, see Figure 9. This characteristic was unexpected with Taguchi sensors, which are nonlinear devices, but it could be expected with linear sensors. The most probable reason why we still experienced this property with our sensor is the linear response of nonlinear systems against small perturbations; a situation relevant for Taguchi sensors.

The measurement conditions to test the impact of bacterium numbers were as follows. Six different bacterium numbers of E. coli were used: $2.5*10^4$, $5*10^4$, $10^5$, $2.5*10^5$, $5*10^5$ and $10^6$. The normalized power density spectra and the binary patterns are shown in Figures 9 and 10, respectively.

**Figure 9.** Variations of the normalized power density spectrum at different bacterium numbers.

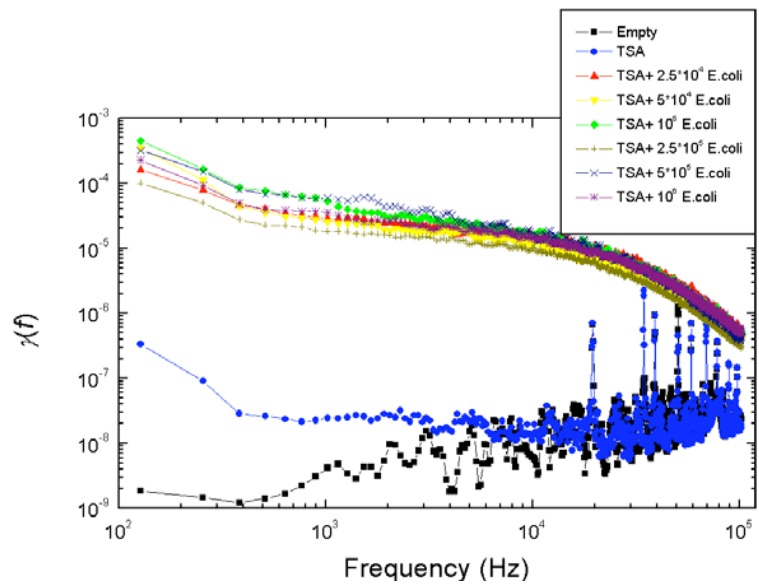

**Figure 10.** Variations of the binary pattern at different bacterium numbers. Bit B5 is not reliable therefore that bit should not be used for pattern recognition, see the Boolean logic in Section 6.

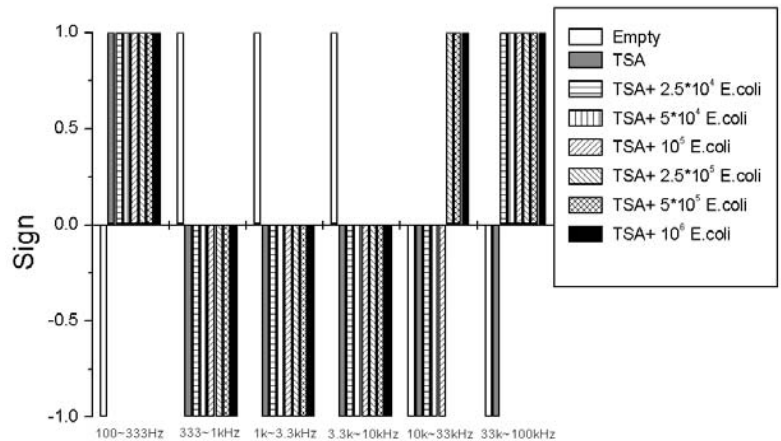

The binary pattern in Figure 10 can also identify three conditions: empty, TSA and TSA+ bacteria (*E. coli*). When the bacterium number decreases from $10^6$ to $2.5*10^4$, the bits remained the same except bit B5 (relevant to sub-band 10k~33kHz). As a consequence, bit B5 should not be used in the Boolean logic for pattern recognition, except perhaps as extra information about the bacterium number. However, the rest of the bits provide sufficient information to identify the different types of samples: empty, TSA and TSA+ bacteria (*E. coli*).

**6. Boolean logic circuit for pattern recognition with ultra-low power need**

The obtained bit patterns are shown in Table 1. As we have shown above bit B5 is not robust against variations in the bacterium number thus we can consider it as an invalid bit unless we want to use that for the indication of bacterium number of *E. coli*. Thus a simple 5-bits Boolean logic circuit driven by pattern bits 1-4 and 6 can identify the different situations. The Boolean logic circuit in Figure 11 has a two-bits output where only 3 of the four possible states are used to display the recognized pattern: *empty*, *TSA*, or *TSA + bacteria* (either *E. coli* or *Anthrax*).

**Table 1.** The logic input of the Boolean logic circuit generated from Figure 10. Zero stands for $\sigma_n = -1$ and 1 stands for $\sigma_n = 1$.

|  | Bit B1 | Bit B2 | Bit B3 | Bit B4 | Bit B5 | Bit B6 |
|---|---|---|---|---|---|---|
| Empty | 0 | 1 | 1 | 1 | N/A | N/A |
| TSA | 1 | 0 | 0 | 0 | N/A | 0 |
| TSA+bacteria | 1 | 0 | 0 | 0 | N/A | 1 |

The output of the Boolean logic at the different situations is shown in Table 2. The output 1,0 is not shown because it is invalid.

**Table 2.** The output of the Boolean logic circuit at the different situations.

|  | Output of Binary logic | |
|---|---|---|
|  | Bit 1 | Bit 2 |
| Empty | 0 | 0 |
| TSA | 1 | 0 |
| TSA+ Bacteria | 1 | 1 |

From Table 1 and Table 2, the following Boolean logic equations can be extracted:

$$\text{Bit 1} = B_1 \cdot \overline{B_2 \cdot B_3 \cdot B_4} \tag{5}$$

$$\text{Bit 2} = B_1 \cdot \overline{B_2 \cdot B_3 \cdot B_4} \cdot B_6 \tag{6}$$

The Boolean logic circuit to realize Equations 5, 6 is shown in Figure 11.

**Figure 11**. The Boolean logic circuit to realize the binary pattern recognition for the sampling-and-hold sensor SP32.

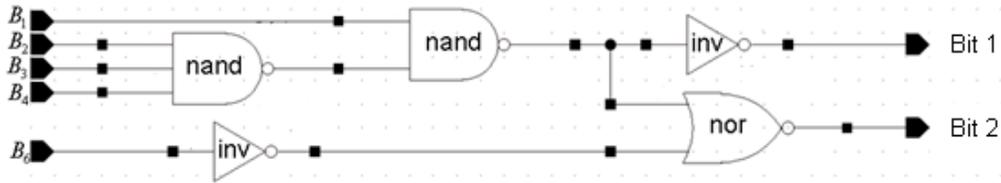

When this Boolean logic circuit is realized by CMOS logic gates, its static power consumption is in the nano/picoWatt regime. In addition, this circuit consumes some extra power for a short time (nanosecond) while it switches when the situation of agents change. Due to the rarity of such events, the power consumption in the Binary logic is basically due to the leakage current of transistors. At practical situations, these powers required by the logic circuitry are negligible.

## 7. Power consumption of the whole sensing system

As a result of avoiding digital computation and using analog processing with the simple logic decisions instead, the main advantage of our system is its simplicity and ultra low power consumption. The analog circuits dominate the power consumption of this system and that is relatively small due to the low-frequency operation (<100kHz). In [24] a more sophisticated analog circuitry for a different sensing approach is shown, including a wireless unit (known to be power hungry), with only 3 microWatt total power dissipation. Thus we can safely claim that the power dissipation of our analog circuitry is in the microWatt range or below. In comparison, a laptop computer based pattern recognizer, which would be able to run the same task, would dissipate around 20-50 Watts.

## 8. Summary

In this work we have reported an exploratory study to generate and test highly distinguishable and robust types of binary patterns from power density spectra obtained at fluctuation-enhanced sensing of bacterial odors. We have shown a way how these binary patterns can be generated by an analog circuitry of ultra-low power consumption and used to drive a Boolean logic based pattern recognizer with negligible power consumption. We demonstrated these findings by single-sensor experiments recognizing bacteria with 100% success rate and zero false alarm rates.

Concerning an important question asked by a Referee about discriminating gram-positive and gram-negative bacteria, the answer is that this method is sensing the odor of the bacteria. If these sets represent some characteristic odor components then classification can be possible. Otherwise, the only way is to teach the system (construct the binary logic) to recognize each specific bacterium.

## Acknowledgment

This work was supported in part by the Army Research Office under contract W911NF-08-C-0031.